# Stabilization of binary NiO-YSZ nanocolloids for direct inkjet deposition

*Massimo Rosa*[1], *Philippe Zielke*[1], *Ragnar Kiebach*[1], *Victor Costa Bassetto*[2], *Andreas Lesch*[2], *Vincenzo Esposito*[1]

[1]DTU Energy, Technical University of Denmark, Risø Campus, Frederiksborgvej 399, 4000, Roskilde, Denmark

[2]Laboratoire d'Electrochimie Physique et Analytique, Ecole Polytechnique Fédérale de Lausanne, EPFL Valais Wallis, Rue de l'Industrie 17, CH-1950 Sion, Switzerland

*corresponding author: masros@dtu.dk

**ABSTRACT**

Water-based inks, containing nanometric NiO and YSZ particles in 66/34 vol. % ratio, are produced by colloidal stabilization of a binary dispersion obtained *via* continuous hydrothermal synthesis at supercritical conditions, *i.e.* 280 bar and 400°C. The method yields single-crystal particles with diameter ≤ 10 nm for both phases in a single-step process, achieving a highly mixed composite. Two different approaches are applied to formulate inks printable with piezoelectric printheads, i.e. an electrostatic and an electrosteric stabilization path. The use of the electrosteric dispersant results in colloids with superior stability > 200 days, more uniform thin films and finely nanostructured porous cermet films with thickness below 500 nm, after reducing NiO to Ni. Particles coarsening to 50-150 nm is obtained at 1000°C, accompanied by a shrinkage of *ca.* 43 % in thickness without the formation of cracks or delamination of the zirconia substrates.

# 1. Introduction

The investigation of functional nanometric ceramics for energy related devices is spreading to several technologies such as photovoltaics [1,2], solid oxide fuel/electrolysis cells [3–7], batteries [8–10] and heat transfer [11]. This is especially due to the unique properties of nanomaterials in terms of performance [3,4,12,13], and their special features in the processing [14,15]. However, exploiting the technological potential of nanostructures can result problematic as it requires control of their synthesis [16,17], agglomeration [18] and of the hazards related with handling nanopowders [19].

Several different methods are available nowadays for the direct synthesis of nanomaterials, such as coprecipitation, sol-gel, and solvo/hydro-thermal synthesis [20]. Among them, the hydrothermal route has been intensively proposed for the synthesis of a wide range of metal oxides nanoparticles with different morphologies [16,17,21]. Due to its versatility, this method has been developed towards continuous flow reactors [17,22], which are now employed in industrial productions [23]. An additional peculiarity of the Continuous Hydrothermal Synthesis (CHS) is the possibility to operate at supercritical conditions, i.e. $scH_2O$: T > 374°C, p > 221 bar. Under such parameters, the dielectric constant and viscosity of water drop to 20 and 30 µPa s respectively, resulting in a suitable environment for producing dispersed nanoparticles with a narrow particle size distributions [24]. The resulting water-based suspensions have a strong potential for application in liquid processing of nanomaterials, completing an all-wet processing strategy. This is particularly suitable for deposition techniques where very low viscosities are required, such as inkjet printing.

Drop-on-demand (DOD) inkjet printing is based on the digitally controlled deposition of droplets. In a typical DOD printer, hundreds to thousands of droplets ranging from picoliters ($10^{-12}$ L) to nanoliters ($10^{-9}$ L) are jetted at high frequencies (thousands of Hz) to print features in the tens of microns range. In the last 20 years, this technique has been investigated for the deposition of metals and ceramic materials due to the advances in the formulation of particle-containing inks [25–28]. Consequently, by inkjet printing it is possible to address the deposition of *ng* of ceramic particles at high spatial resolution [27].

However, to achieve such processing performances, it is crucial to control the colloidal properties of the ink, especially in the case of nanocolloids, which are known to show unique rheology [29,30]. First, particle agglomeration needs to be reduced for keeping the largest aggregates below 1/50$^{th}$ of the nozzle diameter to avoid clogging of the nozzle and printhead [31]. Second, stable and reproducible jetting is only achieved when the ink rheology is optimized for inkjet printing. This has been studied through modelling [32–34] and experimentally [28,34,35], indicating that a specific combination of viscosity and surface tension is required for generating stable droplets [27,35].

In our previous work, the formulation of metal oxide nanocolloids produced by CHS has been shown in the case of Yttria Stabilized Zirconia (YSZ) [36]. Here, we further explored those results by investigating the formulation and printing of a binary ink containing both YSZ and NiO nanoparticles, a typical combination for porous Ni/YSZ electrodes of fuel cells and electrolysers [37]. The starting binary colloid has been produced via CHS at supercritical conditions, using a previously optimized route [38]. This method takes advantage of synthesizing two materials in a single step, producing a highly mixed composite. Despite this advantage, the addition of a second phase to the dispersion introduces an additional challenge for the ink formulation. Stabilizing suspensions with more than one material is indeed difficult due to the need of finding a common pH range where both YSZ and NiO are stable [39–46]. Therefore, avoiding agglomeration and sedimentation of the ink, can be a major challenge compared to the single-material case. To tackle these issues, here we show two different approaches for formulating printable YSZ/NiO nanoinks via electrostatic and electrosteric stabilization. The first approach avoids additives, but is bound to a specific ink composition, while the second provides more flexibility in the ink formulation, but requires the use of dispersants.

## 2. Experimental

2.1 Particle Synthesis and characterization

The continuous hydrothermal reactor at DTU Energy and the synthesis protocol for NiO/YSZ have been described previously [38]. The detailed synthesis of pure YSZ is reported here [36]. For the synthesis of the NiO/YSZ composite herein, a precursor stream of 0.031 mol $L^{-1}$ $ZrO(NO_3)_2$, 0.0054 mol $L^{-1}$ $Y(NO_3)_3$, and 0.14 mol $L^{-1}$ $Ni(NO_3)_2$ with a flow rate of 8.3 mL $min^{-1}$ was mixed with a stream of 1 mol $L^{-1}$ of KOH at 9.8 ml $min^{-1}$. The resulting stream was mixed with supercritical water (flow equivalent to 14.5 mL $min^{-1}$ at room temperature) at 400 °C and 280 bar. The stream of reactants and $scH_2O$ was re-heated to 385°C after mixing, with the aid of a re-heater. The products were eventually cooled down to room temperature and the pressure released. After collection, the particles were concentrated by centrifugation and re-dispersed in clean DI water several times. The dispersion obtained after this step will be referred to as starting dispersion and a pH indicator was used to measure a pH of 10 < pH < 11.

The X-ray diffraction characterization was carried out with a Bruker D8 (Cu Kα radiation) measuring with a 0.05° / 5 s step size / step time. The as-produced particles were analyzed by TEM microscopy, using a Jeol 3000F microscope. The final phase ratio between NiO and YSZ was determined by EDX using a Zeiss Merlin SEM equipped with a Bruker XFlash 6 EDX detector. A few droplets of the starting dispersion were casted onto a microscopy stub covered with carbon tape and let dry. The dried material was analyzed by acquiring EDX spectra from three different areas.

2.2 Ink formulation and characterization

Isopropanol (Sigma Aldrich, analytical grade, IPA), propylene glycol (Sigma, analytical grade, 1,2-PG), 1-propanol (VWR Chemicals, Emsure, ACS Reag. Ph Eur), Dispex A40 (Ciba-BASF, UK), nitric acid (69% VWR Chemicals, Emsure, ACS Reag. Ph Eur) were used as received. For the electrostatic stabilization studies (**Ink1**), IPA, 1,2-PG and the starting dispersion were mixed in a 1:1:1 volume proportion. **Ink1** was mixed for 15 minutes at 35% amplitude in a Sonics Vibra Cell 505 sonicator in the "Cup horn" configuration with 5 s pulses alternated with 5 s at rest. In the case of electrosteric stabilization (**Ink2**), the dispersion was sonicated with a Hielscher UP200ST sonicator using a tip configuration and applying 120 s bursts at 50% amplitude upon each addition of Dispex A40. The ink bottle was kept immersed in water to avoid overheating. Prior to that, the pH was lowered to 8.5 using $HNO_3$ and the pH was confirmed a pHenomenal pH 1000 L pH-meter. The particle size distribution was measured with dynamic light scattering (DLS) and laser diffraction to cover the whole range of particles and agglomerates. A Malvern Zetasizer was employed for the DLS measurements and the samples were prepared by diluting the starting dispersion to 0.1% wt using the same solvent composition and adjusting the pH to desired values by adding $HNO_3$. The pH values waere measured with a Metrohm 744 pH-meter. Laser diffraction measurements of the particle size distribution (PSD) were carried out with a Beckman Coulter, LS 13320. The viscosities of the inks were measured using a SV-1 A series viscometer (A&D Instruments Limited) and an Anton Paar MCR 302 rheometer in rotational mode, considering the viscosity value at a shear rate of 1000 $s^{-1}$. All the measurements were carried out at room temperature. The surface tension was measured with a bubble pressure tensiometer (BP50,

KRÜSS) and a drop shape analyzer DSA-30 (Krüss) in pendant droplet geometry. The printability parameter Z was calculated for comparing the properties of the inks, as reported in previous publications [26]. The dimensionless parameter Z was proposed as a good descriptor for taking into account the most important fluid properties for the jetting process, and has the form: $Z = (\rho \cdot \sigma \cdot a)^{1/2}/\eta$ [32], where $\sigma$ the surface tension, $\rho$ is the density, $\eta$ the viscosity, and $a$ is the characteristic length, typically associated with the nozzle diameter. Different $Z$ ranges are reported in literature where optimal jetting has been observed, such as 1-10 [27], 4-14 [35], and 4-20 [36]. Solid loading of the dispersions was measured by DTA/TG performed on the starting dispersion using a STA 409 PG (Netzsch) at a constant rate of 10 K min$^{-1}$ until a maximum temperature of 500°C under air.

2.3 Inkjet printing

Inks were jetted and printed using two printers: a Pixdro LP50 and a Fujifilm Dimatix DMP-2850 both equipped with the same DMC disposable piezoelectric printheads from Dimatix. These printheads have 16 nozzles of 21.5 μm in diameter for jetting droplets with a nominal volume of 10 pL. According to the printhead specification, the best performances are achieved using inks with a viscosity of 10 mPa s and a surface tension of 30 mN m$^{-1}$. For **Ink1**, the waveform actuating the piezoelectric elements responsible for the droplet ejection was optimized starting from a standard trapezoidal pulse. This waveform was characterized by a jetting segment of 4–7 μs, a dwell segment of 10 μs at 40-60V and fall time 5 μs. A jetting frequency of 1000 Hz was used for all the tests. A splat spacing of 63 μm was employed for printing the multi-layered films. For **Ink2**, a sigmoidal waveform was applied with a total pulse duration of ca. 11 μs and maximum voltage between 13 and 18 V. In this waveform, the jetting pulse is preceded by a filling pulse where the piezo element is moved in the opposite direction with a potential of -10 V. Multi-layers were printed with a 40 μm droplet spacing. All the samples were printed onto in-house produced 8%mol YSZ (TZ-8Y powder from Tosoh) pre-sintered substrates. Before printing, **Ink1** was filtered using a 700 nm mesh syringe filter and **Ink2** was filtered using a 1.3 μm mesh syringe filter. The droplet radius and eventually their volume was determined by image analysis.

2.4 Sintering and thermochemical reduction

The printed layers were sintered in air using the following thermal cycle: RT – 600°C at 15°C h$^{-1}$; dwell for 2h; 600°C – 800°C, 1000°C or 1295°C at 60°C h$^{-1}$; dwell for 6h; cooling down at 100°C h$^{-1}$. The sintered materials were treated overnight at 800°C with a $H_2/N_2$ 9% vol flow for the reduction of NiO to Ni. The resulting microstructures were analyzed using a Zeiss Merlin SEM and a FEI Teneo SEM. Porosity was estimated by image analysis.

3. **Results and discussion**

The processing of NiO/YSZ began with the binary material synthesis of the two phases by continuous hydrothermal synthesis. Fig.1A shows a sketch of the continuous hydrothermal reactor used with the selected configuration for the streams of reactants and $H_2O$ at supercritical conditions (sc$H_2O$).

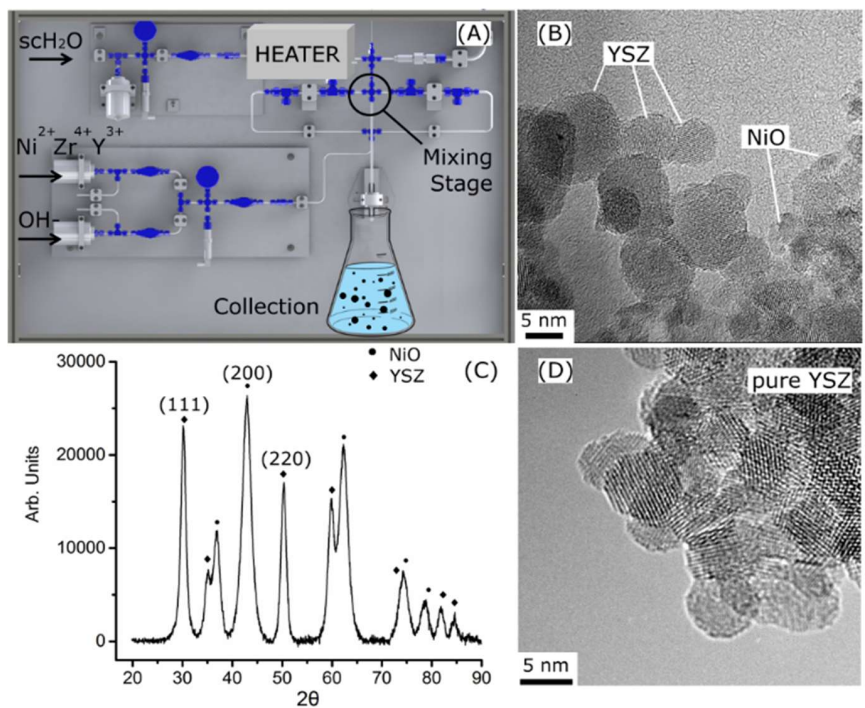

*Fig. 1: (A) sketch of the continuous flow reactor, (B) TEM images of the NiO/YSZ nanoparticles, (C) diffractogram of the synthesized material, and (D) TEM images of YSZ nanoparticles synthesized independently.*

Ni, Zr and Y nitrates were fed into the system and pre-mixed with a KOH solution in a T-shape mixer. This stream was eventually mixed in a co-flow geometry with the $scH_2O$ stream, and the resulting dispersion was cooled down and collected after releasing the pressure.

The morphology of the as-produced nanoparticles was analysed by TEM microscopy (fig. 1B). The particle size of the NiO/YSZ composite was comprised between 4 and 10 nm, with NiO forming 40% smaller particles in average, compared to YSZ. Particles of the two materials were identified in the micrograph by comparing the measured lattice spacing with the position of the diffraction peaks. Three peaks were chosen: (111) and (220) for YSZ and (200) for NiO, due to their distance from other reflections. By measuring the fringe spacing of the particles in the TEM images, it was possible to associate them to the aforementioned peaks and identify YSZ or NiO particles. In fig. 1B, three YSZ particles were highlighted with a spherical morphology and a diameter of 4-10 nm. From the same micrograph, it was possible to distinguish two NiO particles, labelled, which were more elongated than YSZ with a diameter of ca. 3-5 nm. The other particles in fig. 1B showed lattice spacings corresponding to reflections which are overlapped between YSZ and NiO, thus they cannot be uniquely identified. The TEM micrograph highlighted the crystallinity of the produced materials that was confirmed by the result of the XRD measurement, where the peaks for both phases, YSZ and NiO, can be identified (fig. 1C). The shape of the XRD peaks indicated the presence of nanosized crystalline domains, which, according to the Sherrer equation, measured 7 nm for YSZ and 4 nm and NiO. This result correlated well with the data obtained from microscopy and confirmed that the synthesized nanoparticles are made of individual crystallites. The YSZ obtained in the binary material synthesis (fig. 1B) was compared to the pure YSZ (fig. 1D). The size and morphology of YSZ in the two syntheses were similar, suggesting a negligible interaction with NiO in the nucleation and growth mechanism. Finally, an elemental analysis was carried out by EDX on the dried powders for determining the discrepancy between the composition of the precursor solution and the final material. Precursors solutions were prepared to have an elemental molarity ratio of Ni:Zr:Y of 0.79:0.18:0.03, corresponding to a final desired volume ratio of NiO/YSZ = 0.675/0.325 vol. %. This ratio is representative for typical compositions used for the fabrication of functional Ni/YSZ cermets [37]. The EDX analysis showed a composition of 0.77 : 0.20 : 0.03 for Ni : Zr : Y in the final product, indicating a slightly higher conversion of Zr than for Ni. The yields of the two reactions were thus similar and allowed to obtain a final composite with a composition close to the desired one, i.e. NiO/YSZ = 0.66/0.34 vol. % ratio (see Fig. A1 and Table A1 in the Appendix A).

For synthesizing NiO particles with size < 10 nm, KOH needed to be added at the mixing stage, where it acted as mineralizer and reduced the NiO particle diameter [38]. Consequently, the produced dispersion was strongly basic, differently from the synthesis of pure YSZ, where the starting dispersion for the ink formulation showed a pH of 3. For NiO/YSZ, the starting dispersion had a pH > 10 with a strong impact on the particle size distribution (PSD). In particular, fig. 2 shows that, in the pure YSZ dispersion (B), only a small fraction of particles formed small agglomerates < 200 nm. On the other hand, the NiO/YSZ composite contained a large fraction of agglomerates with size > 1 μm that exceeded the maximum printable size of 400 nm. This result indicates that stabilizing the colloid is a crucial step to achieve a printable ink.

Two different stabilization strategies were investigated, as shown in fig. 2. The first one only relied on electrostatic variations of the particles chemical environment, *i.e.* changing pH and supernatant polarity (fig. 2A). In the second strategy, a commercial dispersant, Dispex A40, was used to give a sterical contribution to the colloidal stabilization. Dispex A40 was chosen as a dispersant due to its effectiveness in dispersing metal oxide particles in aqueous media under basic pH conditions [47]. For the two strategies, the results were expected to be complimentary. The purely electrostatic strategy avoids the use of organic additives, which lower the green density of the material, but requires a careful control of the pH depending on the solvent polarity. Conversely, the electrosteric strategy (fig. 2C) introduces a polymeric dispersant, but is less sensitive to the addition of co-solvents for adjusting the ink properties.

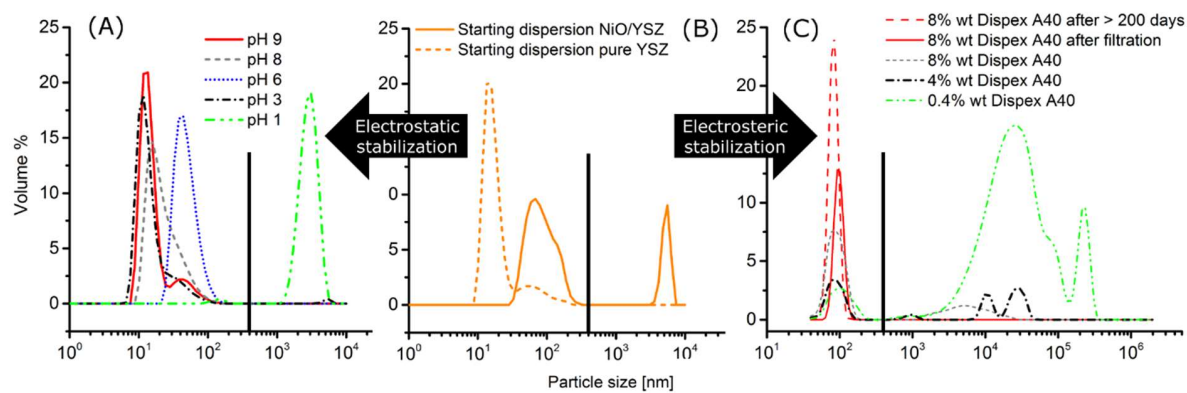

*Fig. 2: PSD of the NiO/YSZ dispersion after addition of IPA and 1,2-PG at different pH (A), PSD of the aqueous starting dispersion of NiO/YSZ at pH 10-11 (B solid line). Pure YSZ dispersion is shown for comparison (B dotted line). PSD of the NiO/YSZ dispersion upon the addition of an increasing amount of the dispersant Dispex A40 (C). The black vertical line in all plots indicates the maximum agglomerate size for avoiding nozzle clogging, i.e. 400 nm.*

For reaching a printable **Ink1** via electrostatic stabilization, the influence of pH on the PSD was investigated. PSD data were collected using the exact solvent composition of the final ink, a crucial parameter for electrostatic stabilization. Pure water as solvent has a too low viscosity and too high surface tension. Therefore, prior to the PSD measurements, IPA was added to the starting dispersion for decreasing the surface tension to 30 mN m$^{-1}$ (see table 1). However, due to the low boiling point of IPA (B.P. 78 ° C), 1,2-PG (B. P. 188 ° C) was also added to reduce the ink evaporation rate, and the related risk of nozzle clogging. Fig. 2A reports the PSD curves of the mixed dispersions at different pH values. At extremely basic and acidic conditions, pH 1 and pH > 11, particles agglomerated in a few seconds, as it can be seen for the curve at pH 1. In the case of pH 11, the agglomeration took place too fast and it was not possible achieving a reliable particle size measurement (no curve for pH 11 reported in fig. 2A). Differently, for pH values between 3 and 9, the measured PSD approached values comparable to what was observed by microscopy. Indeed, the maximum of the curves was comprised between 10 and 100 nm, indicating a strong reduction of the agglomerates close to the primary particle size. In this solvent system, a large pH window between 3 and 9 existed, where the resulting PSD respected the size limit for inkjet printing, without the addition of an organic dispersant.

In the electrosteric strategy to reach **Ink2**, different amounts of Dispex A40, the ammonium salt of a polyacrilic acid, were added directly to the starting dispersion, as it can be seen in fig. 1C. The pH of the starting dispersion was adjusted to 8.5 in order to reach the pH range where Dispex A40 is most effective, according to what suggested by the producer. Fig. 1C shows a first large increase in the agglomerate size upon addition of 0.4% wt of Dispex A40 due to bridging. Bridging is a result of the polymeric chains binding to more than one particle and thus forming agglomerates [18]. This indicated that 0.4% wt of dispersant was not enough to cover the particles' surface completely. By adding 4% wt of dispersant, the number of agglomerates due to bridging reduced notably, as most of the volume fraction of the material lay below 400 nm. However, the resulting dispersion was still not printable since agglomerates with a diameter of 1 μm and > 10 μm were still present. For this reason, the amount of dispersant was increased gradually until a plateau was reached at 8% wt, where the agglomerates formed a broad and low intensity peak between 1 – 100 nm. Filtering of the dispersion through a 0.7 μm filter was done to secure the elimination of the few larger agglomerates that had remained. The curve in fig. 2C confirmed the absence of agglomerates after filtering. Thermogravimetric analyses on the dispersion before and after filtration showed a solid loading variation of only 0.12% wt, from 6.36 %wt to 6.24 %wt (see Fig. A2 in Appendix A). This result indicates that only a minor part of the material was agglomerated and removed by filtration. Moreover, the PSD was measured on the same dispersion after >200 days on the shelf, showing still a printable size distribution and demonstrating a very good long-term colloidal stability. Thus, this approach resulted in a highly dispersed colloid maintaining a printable PSD for more than 6 months.

The dispersions stabilized by these two strategies were then investigated for their direct use for inkjet printing. Among the colloids stabilized electrostatically, the dispersion at pH 9 was selected as it fulfilled all printing requirements. Viscosity, surface tension and density were measured and the obtained values are reported in table 1. The σ and η values fit the printhead specifications and the printability parameter Z lay in the optimal range between 1 and 10. This ink will be referred to as **Ink1**.

Differently, a surface tension of 60 mN m$^{-1}$ was measured for the colloid stabilized by the second approach, hence, 30% wt of 1-propanol was added for decreasing σ to 25 mN m$^{-1}$. The resulting ink showed a lower viscosity than **Ink1**, with a Z value of 11, see table 1. This was higher than the empirical limit of 10, yet was still in the range of good printability observed in a previous study on nanometric YSZ (4 < Z < 20). This second ink, with the addition of the dispersant, is here referred as **Ink2**.

*Table 1: properties of the inks formulated using the two different stabilization approaches.*

| Ink name | Viscosity mPa s | Surface tension mN m$^{-1}$ | Density g cm$^{-3}$ | Z | Solid loading %wt |
|---|---|---|---|---|---|
| Ink1 | 6 | 33 | 0.95 | 4.3 | 2.2 |
| Ink2 | 2 | 25 | 0.93 | 11.2 | 4.7 |

The experimental printability of the two inks was studied by depositing layers and patterns onto a pre-sintered 8YSZ substrate. The waveform for actuating the piezoelectric elements of the printhead was optimized for both inks. In the case of **Ink1**, a sigmoidal waveform was used to achieve a stable jetting of single droplets, see fig. 3A. For **Ink2**, a trapezoidal waveform was employed for achieving single, round-shaped droplets (fig. 3B). Therefore, both inks could be jetted successfully forming stable droplets with a volume comprised between 7-9 pL.

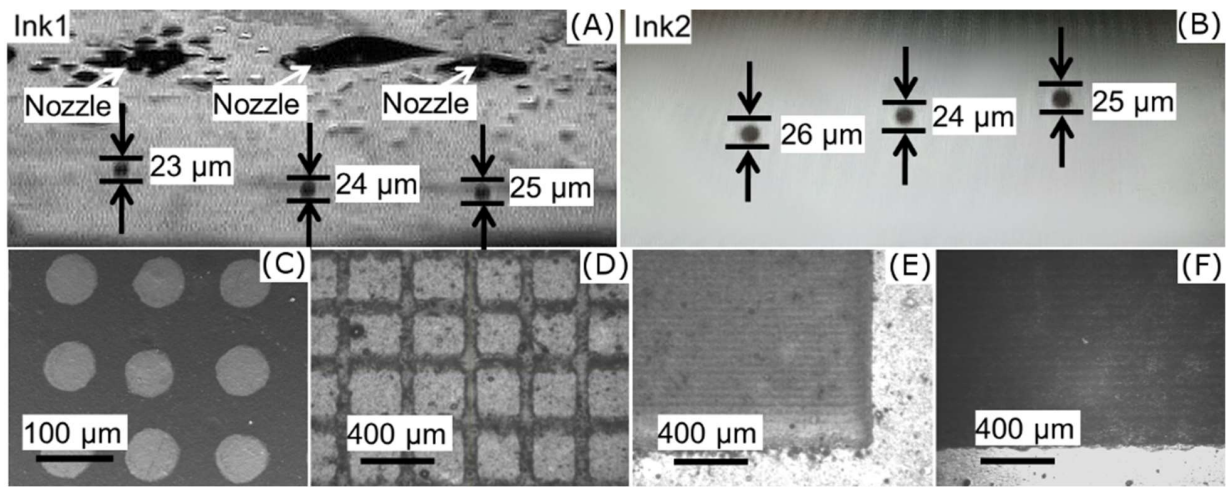

*Fig. 3: jetting of the inks formulated in the two strategies, **Ink1** (A), **Ink2** (B). Separated splats (C) and fine grid (5 layers, D) printed with **Ink1**. Continuous films printed onto dense YSZ with **Ink1** (E) and **Ink2** (F).*

Using the optimized jetting conditions, the behaviour of the inks during printing was investigated. Separated splats were printed to define the spatial resolution of the process. A splat diameter of ca. 50 μm was measured for **Ink1** (fig. 3C), which represents the smallest printable feature under these conditions. Noteworthy, the splats showed a remarkably uniform distribution of the particles after drying, *i.e.* a negligible coffee stain effect. These printing capabilities were demonstrated by depositing complex patterns, such as the grid reported in fig. 3D. Finally, continuous films up to 1.5x1.5 cm$^2$ were printed (see Fig. A3 in the Appendix A) increasing the areal solid loading by printing multiple layers up to values of 0.21 mg cm$^{-2}$ for **Ink1** (fig. 3E) and 0.10 mg cm$^{-2}$ for **Ink2** (fig. 3F). These values refer to the final amount of solid particles deposited.

The continuous films were used to study the material behaviour during the sintering and thermochemical reduction of NiO. Films printed with both inks were sintered in air at 1000°C and SEM analyses were carried out to observe the resulting microstructures (fig. 4).

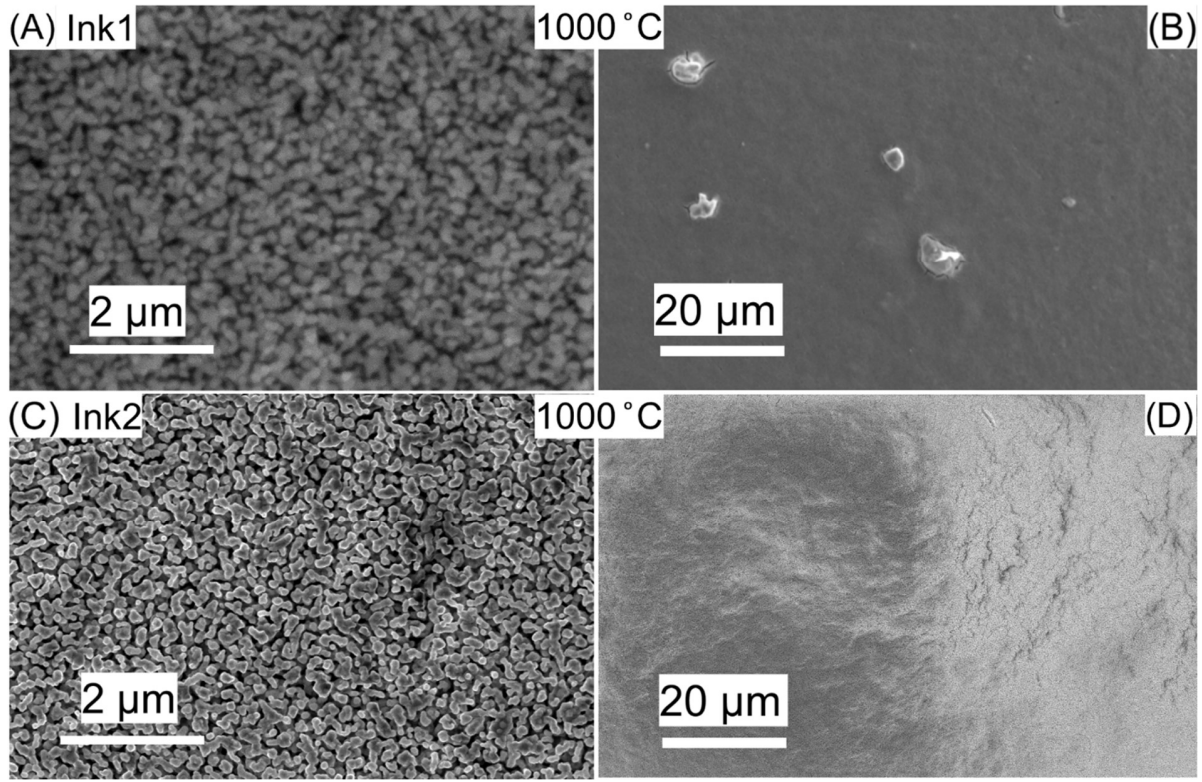

*Fig. 4: microstructure after sintering at 1000°C of films printed with **Ink1** (A,B) and **Ink2** (C,D). Low magnification images of the films: agglomerates with **Ink1** (C) and uniform film with **Ink2** (D).*

Fig. 4A and fig. 4C compare the microstructure at high magnification of the materials printed with **Ink1** and **Ink2**. They were both characterized by a very open structure with a continuous network of pores and interconnected particles with size between 50 and 150 nm. Therefore, at 1000° C, there was already a pronounced grain growth compared to the untreated particles ($\leq$ 10 nm). At this magnification, no big differences could be identified between materials printed with the two inks. Such a result was unexpected as the **Ink2** contained an organic dispersant that lowered the green density of the deposited material. Densification is probably prevented by the constrain of the pre-sintered substrate, regardless of the ink composition. On the other hand, differences could be observed on a larger scale, represented here by fig. 4B and fig. 4D. In particular, large agglomerates were present and distributed on the whole surface of the films printed with **Ink1**, when the areal solid loading is $\geq 0.13$ mg cm$^{-2}$. These agglomerates have large dimensions, comprised between 1 and 10 µm, and formed necessarily after printing. Indeed, the PSD of the ink did not indicate the presence of such agglomerates, as previously discussed. Moreover, discontinuous jetting and clogging of the printhead would be expected with such large particles, but was not observed. After deposition, the particles possibly agglomerated during drying of the ink, when the supernatant polarity gradually changed due to the different boiling points of the solvents, accompanied by a pH increase.

Differently, **Ink2** formed crack-free films on a centimetre-size scale after sintering at 1000 °C, see fig. 4D. In this case, the presence of the organic dispersant covering the particle surfaces prevented agglomeration, also during the evaporation of the supernatant. Therefore, the electro-steric stabilization represents a reliable option for the fabrication of high quality films with a uniform microstructure.

After comparing the two strategies, samples printed with **Ink2** were chosen for their better microstructure to investigate in detail the last processing step, *i.e.* thermal treatments. In particular, the microstructural analyses were carried out after the reduction of NiO to metallic Ni in order to correlate the sintering temperature to the microstructure of the final cermet. Films produced by printing **Ink2** were sintered at three different temperatures, 800, 1000 and 1295°C in air and were subsequently reduced in a $N_2/H_2$ atmosphere at 800°C for 18h. Fig. 5A-F reports top-view and cross section micrographs for each temperature.

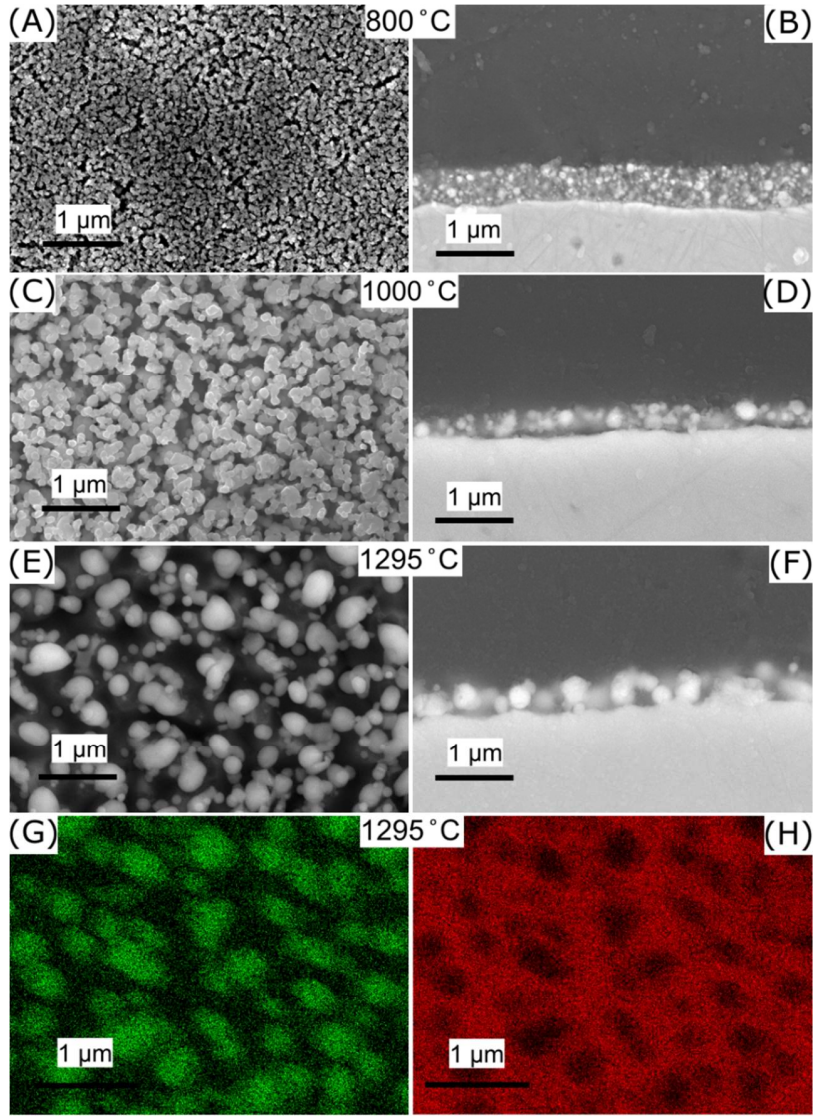

*Fig. 5: Top view and cross section of the microstructures obtained after sintering at 800°C (A,B), 1000°C (C,D) and 1295°C (E,F), each followed by reduction under $H_2/N_2$ at 800°C. EDX mapping of the sample sintered at 1295°C, (G) Ni signal, (H) Zr signal.*

The images show an evident coarsening of particles and porosity with temperature. At 800°C, the nanosize of the particles was still preserved and the porosity formed an interconnected network with a width < 100 nm and *ca.* 35% vol porosity (see Fig. A4 in Appendix A). The cross-section image highlighted a continuous and uniform cermet film having a thickness comprised between 450-600 nm. At 1000°C, the microstructure after reduction of NiO (fig. 5C) had larger pores compared to the oxidized material in fig. 4C, due to the loss of oxygen. The homogeneous increase of the porosity indicated a homogeneous distribution of the NiO particles in the deposited material. The mass diffusion and particles coarsening led to a shrinkage of the film with a subsequent reduction in thickness to 250 – 350 nm (fig. 5D). This corresponds to a large shrinkage in the *z*-direction of ca. 43%, while no dimensional variation took place in the x-y directions, due to constrain given by the substrate [48–50]. Noteworthy, the stresses generated during this thermal treatment did not form cracks in the printed films. A further increase in the sintering temperature to 1295°C led on the one hand to a particle growth in the submicron range, fig. 5E. In this case, many particles were disconnected from each other, and the substrate was easily visible. EDX mapping (fig. 5G, 5H) allowed interpreting the obtained microstructure. In particular, the elemental analysis demonstrated that, in addition to the particle growth, only Ni particles were left on top of the substrate, while the printed YSZ inter-diffused in the substrate. Therefore, despite being 1295°C a typical sintering temperature for NiO/YSZ components [26], this temperature was too high for treating the nanostructured thin films. In conclusion, to maintain a nanosized structure of the material, the temperature should be kept as low as 800°C, while should not exceed 1000°C to prevent the diffusion of YSZ.

4. **Conclusions**

A binary NiO/YSZ composite with volume ratio 66/34 was synthesized in a single step process through CHS. Despite the formation of the two materials took place simultaneously, the final composition of the composite is close to the targeted one, with only 2.5% vol less of NiO than the stoichiometry of the precursor solution. Well-mixed particles were produced with a diameter of 4-10 nm for YSZ, and 3-5 nm for NiO. Printable inks were formulated by stabilizing the synthesis product with two approaches, electrostatic and electrosteric stabilization. The ink produced by electrosteric stabilization resulted in more uniform films and showed an ink stability for > 200 days on the shelf. Therefore, the application of an electrosteric stabilization was more suitable for the deposition of NiO/YSZ nanostructured layers.

After sintering at 800°C and reduction of NiO to Ni, films of 450-600 nm thickness were obtained. The material underwent shrinkage of 43% along the cross-section at 1000°C, associated with the coarsening of the particles to a diameter of 50-150 nm. Further increasing the sintering temperature to 1295°C led to the diffusion of the printed YSZ in the substrate. These results highlighted the high mass diffusion of these nanomaterials, and indicated that for keeping a nanostructured material, sintering temperatures as low as 800°C had to be used.

**Acknowledgements**

This project has partially received funding from the Fuel Cells and Hydrogen 2 Joint Undertaking under grant agreement No 700266. This Joint Undertaking receives support from the European Union's Horizon 2020 research and innovation program and Hydrogen Europe and N.ERGHY. The authors also acknowledge funding from Swiss National Science Foundation through the Ambizione Energy Project No. 154297.

# Appendix A. Supporting information

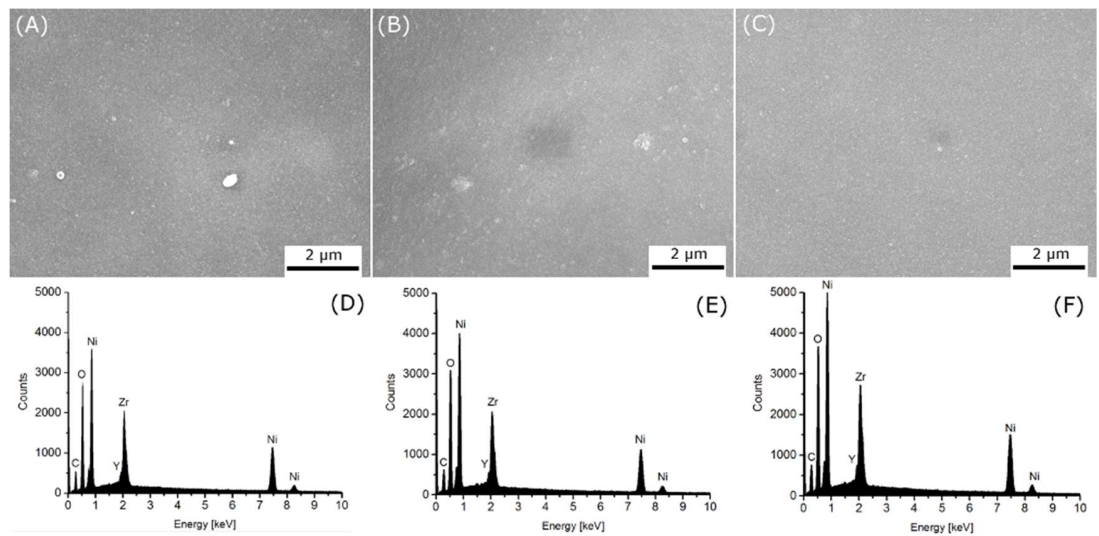

*Fig. A1: (A), (B) and (C), SEM images of the areas analysed for the elemental analysis. (D), (E) and (F), EDX areal spectra of the areas (A), (B) and (C) respectively.*

*Table A2: atomic percentages of Ni, Zr and Y detected in the three areas and corresponding volume percentages of NiO and YSZ.*

|            | Area 1 | Area 2 | Area 3 | Average | Std. Dev. |
|------------|--------|--------|--------|---------|-----------|
| **Ni at%** | 0.77   | 0.77   | 0.77   | 0.77    | 0.0024    |
| **Zr at%** | 0.20   | 0.20   | 0.20   | 0.20    | 0.0020    |
| **Y at%**  | 0.03   | 0.03   | 0.03   | 0.03    | 0.00056   |
| **NiO vol%** | 0.67 | 0.66   | 0.66   | 0.66    | 0.0030    |
| **YSZ vol%** | 0.33 | 0.34   | 0.34   | 0.34    | 0.0030    |

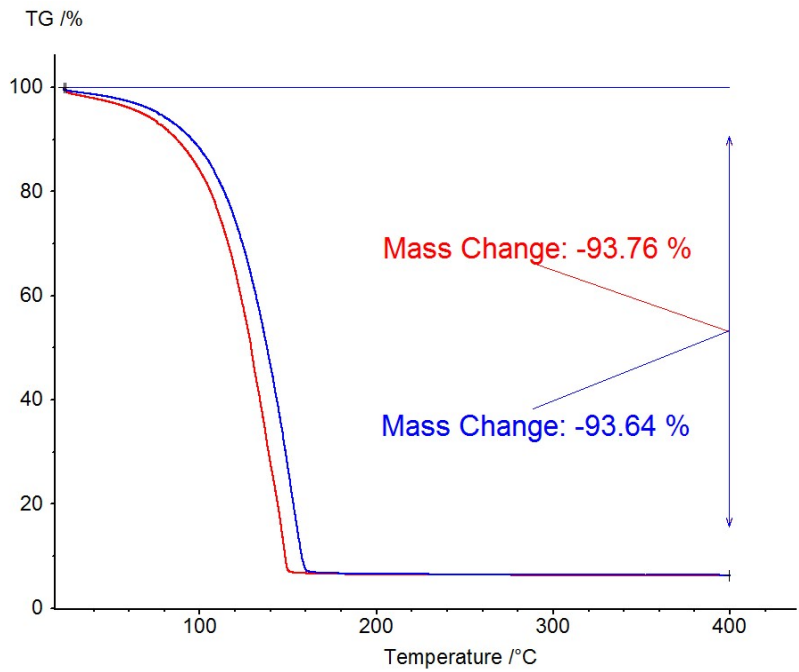

*Fig. A2: thermogravimetric analysis of the **Ink2** before (blue curve) and after (red curve) filtration with a 700 μm syringe filter.*

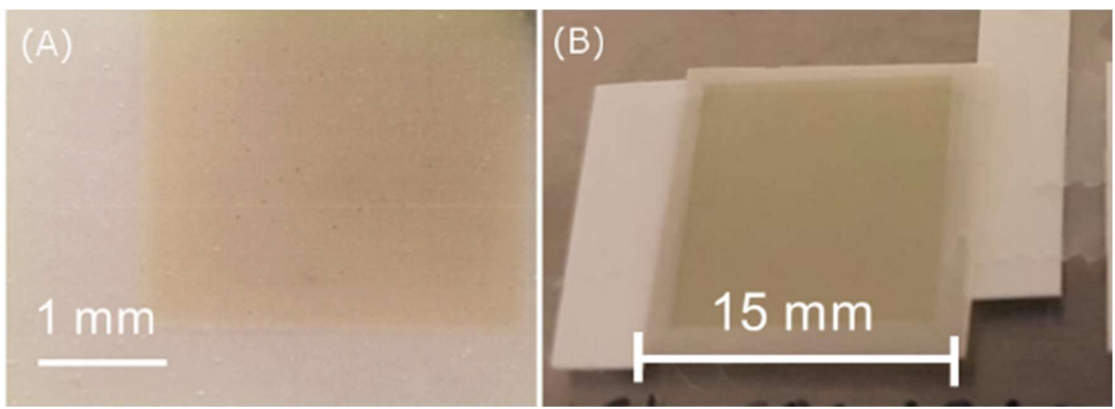

*Fig. A3: (A) optical microscope image of a film printed with **Ink2** after sintering at 800° C. (B) Image of a green film printed with **Ink1**.*

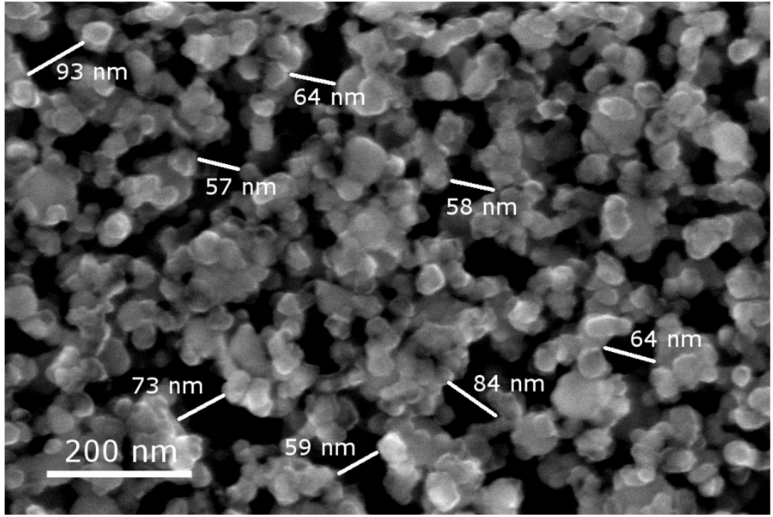

*Fig.A4: high magnification SEM image of the film printed with **Ink2** after sintering at 800° C and reduction in $N_2/H_2$.*